\documentclass[journal,twoside,web]{ieeecolor}
\usepackage{generic}
\usepackage{cite}
\usepackage{amssymb,amsfonts}
\usepackage{algorithmic}
\usepackage{graphicx}
\usepackage{textcomp}
\usepackage{subfigure}
\usepackage{multirow}
\usepackage{graphicx}
\usepackage{multirow}
\usepackage{amsfonts}
\usepackage{amssymb,}
\DeclareMathAlphabet\mathchorus     {T1}{qzc} {m} {n}
\usepackage{dsfont}
\usepackage[utf8]{inputenc}

\usepackage{array} 
\usepackage{mathtools}
\usepackage{fontenc}
\usepackage{tikz}
\usepackage{tikz}
\usepackage[linesnumbered,ruled,vlined]{algorithm2e}

\usepackage{pgfplots}

\definecolor{mygreen}{HTML}{03C03C}
\definecolor{mypink}{HTML}{008000}
\definecolor{Mycolor}{HTML}{FFC40C}
\usepackage[linesnumbered, ruled]{algorithm2e}
\makeatletter
\usepackage[pscoord]{eso-pic}
\newcommand{\placetextbox}[3]{
	\setbox0=\hbox{#3}
	\AddToShipoutPictureFG{ \put(\LenToUnit{#1\paperwidth},\LenToUnit{#2\paperheight}){\vtop{{\null}\makebox[0pt][l]{#3}}}
	}
}
\placetextbox{.172}{0.055}{\scriptsize This article has been accepted in IEEE Transactions on Industrial Informatics Journal © 2025 IEEE. Personal use of this material is permitted.}
\placetextbox{.082}{0.045}{\scriptsize Permission from IEEE must be obtained for all other uses, in any current or future media, including reprinting/republishing this material for advertising or promotional purposes,}
\placetextbox{.152}{0.035}{\scriptsize creating new collective works, for resale or redistribution to servers or lists, or reuse of any copyrighted component of this work in other works.}
\placetextbox{.36}{0.025}{\scriptsize This work is freely available for survey and citation.}

\@ifundefined{showcaptionsetup}{}{%
 \PassOptionsToPackage{caption=false}{subfig}}
\usepackage{subfig}
\makeatother

\def\BibTeX{{\rm B\kern-.05em{\sc i\kern-.025em b}\kern-.08em
    T\kern-.1667em\lower.7ex\hbox{E}\kern-.125emX}}
\renewcommand\thesection{\arabic{section}}

\renewcommand{\thesubsubsection}{\Alph{subsubsection}}

\begin{document}
\title{A Self-Healing and Fault-Tolerant Cloud-based Digital Twin Processing Management Model}
%\author{First A. Author, \IEEEmembership{Fellow, IEEE}, Second B. Author, and Third C. Author, Jr., \IEEEmembership{Member, IEEE}
\author{Deepika Saxena, \IEEEmembership{ Member, IEEE} and Ashutosh Kumar Singh, \IEEEmembership{Senior Member, IEEE}
\thanks{Deepika Saxena is with the Division of Information Systems, University of Aizu, Japan and also with Department of Computer Science, the University of Economics and Human Sciences, 01-043 Warsaw, Poland.  (Email: deepika@u-aizu.ac.jp, 13deepikasaxena@gmail.com).} 
\thanks{Ashutosh Kumar Singh is with the Department of Computer Science and Engineering, Indian Institute of Information Technology Bhopal, Bhopal 462003, India, and also with Department of Computer Science, the University of Economics and Human Sciences, 01-043 Warsaw, Poland. (E-mail: ashutosh@iiitbhopal.ac.in).}}
\maketitle

\begin{abstract}
Digital twins, integral to cloud platforms, bridge physical and virtual worlds, fostering collaboration among stakeholders in manufacturing and processing. However, the cloud platforms face challenges like service outages, vulnerabilities, and resource contention, hindering critical digital twin application development. The existing research works have limited focus on reliability and fault tolerance in digital twin processing. In this context, this paper proposed a novel Self-healing and Fault-tolerant cloud-based Digital Twin processing Management (SF-DTM) model. It employs collaborative digital twin tasks resource requirement estimation unit which utilizes newly devised  Federated learning with cosine Similarity integration (SimiFed). Further, SF-DTM incorporates a self-healing fault-tolerance strategy employing a frequent sequence fault-prone pattern analytics unit for deciding the most admissible VM allocation. The implementation and  evaluation of SF-DTM model using real traces demonstrates its effectiveness and resilience, revealing improved availability, higher Mean Time Between Failure (MTBF), and lower Mean Time To Repair (MTTR) compared with non-SF-DTM approaches, enhancing collaborative DT application management. SF-DTM  improved the services availability up to 13.2\%  over  non-SF-DTM-based DT processing.
\end{abstract}


\section{Introduction}
\label{sec:introduction}
\begin{figure} [!htbp]
    \centering
    \includegraphics[width=0.9\linewidth]{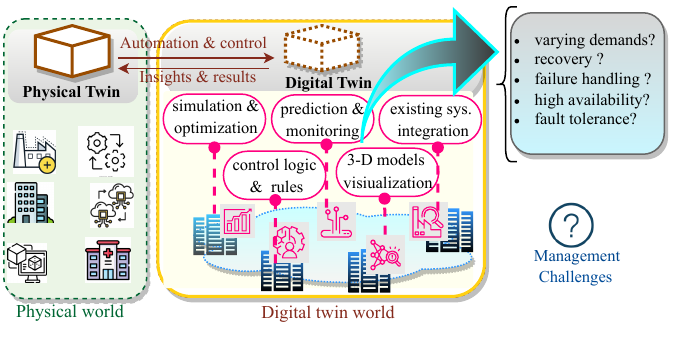}
    \caption{Cloud-driven collaborative DT and challenges}
    \label{fig:enter-label}
\end{figure}

\IEEEPARstart{D}{igital } twin (DT) connects the physical and digital world by integrating Internet-of-Things (IoT) \cite{xu2023survey}, machine learning \cite{ren2022machine}, robotics and virtual reality at its foundation \cite{li2021semantic}. Cloud platforms augment the creation and management of robust DT solutions, offering flexibility, and computational power required for managing and analyzing the vast amounts of data generated by them with high scalability for diverse customer environments \cite{dang2021cloud, wang2023cooperative, jeon2023intelligent}.  Industries such as manufacturing, healthcare, smart cities, aerospace, and defense are utilizing this transformation, shifting from a physical-centric to a digital twin-centric paradigm, facilitated through cloud platforms \cite{bellavista2021application}. As illustrated in Fig. \ref{fig:enter-label}, the cloud-based DTs enable collaborative development environments, empowering stakeholders to utilize services like simulation, optimization, prediction, monitoring, control logic, system integration, and visualization.
These services can be customized to specific business needs \cite{dang2021cloud}. For instance, the cloud platforms facilitate collaboration among multiple stakeholders by enhancing grid reliability and resource optimization towards sustainable energy solutions. However, despite the benefits, cloud based DT faces challenges such as vulnerability to service outages, security threats (due to multi-tenant shared computing), and resource congestion due to concurrent usage and dynamic fluctuations in resource demands. These issues disrupt collaborative DT applications development, leading to productivity loss and missed deadlines. To address these entangled critical challenges, there is a high necessity of a proactive computing resource reservation and planning with resilient collaborative application-focused cloud management design. Fault tolerance strategies, resource management, and contingency plans are crucial for ensuring reliability and performance in cloud-based  collaborative DT development and management.

\subsection{Related Work}
The existing significant research works have attempted to address the issues of adaptive and dynamic resource allocation for cloud supported DT development includes \cite{song2022distributed, jeon2023intelligent,peng2022distributed, lu2020communication,yang2022optimizing, shen2024cloud}. A blockchain-based distributed resource allocation scheme is proposed in \cite{song2022distributed} to enhance the Quality of Service (QoS) for Virtual Reality-embedded DT  services in the manufacturing industry's Industrial IoT landscape. Jeon et al. \cite{jeon2023intelligent} have addressed the problem of  resource scaling  for efficient consumer electronics DT simulation in high-performance cloud computing by optimizing resource utilization and performance through predictive scaling. Also, a two-stage incentive mechanism is presented in \cite{peng2022distributed} for optimizing computation offloading and resource allocation in DT empowered edge networks, ensuring privacy and information security of mobile devices. Federated learning (FL) is used in \cite{lu2020communication} for optimizing communication efficiency and reducing energy costs for integrating DT with edge networks. Further,  FL and asynchronous FL scheme is utilized in \cite{yang2022optimizing} to optimize model construction and device selection, enhancing real-time processing and decision-making in Industry 4.0.
 Shen et al. \cite{shen2024cloud} proposed a cloud-edge collaboration framework utilizing real-virtual collaborative process tracking  to efficiently generate process DTs for remote task supervision, optimizing cost-effectiveness while ensuring high traceability.

Recently some pioneering fault-tolerant cloud resource management works have been proposed including \textit{Failure Aware and Energy-Efficient} (FAEE) VM placement scheme \cite{sharma2019failure} which applied exponential smoothing for failure prediction, Deep neural network-based failure \textit{Prediction for Energy-aware Fault-tolerant Scheduling} (PEFS) scheme \cite{marahatta2020pefs},  VM \textit{Significance Ranking and Resource Estimation based High availability Management} (SRE-HM) Model \cite{saxena2022high} which applied LSTM-based fault estimation, and \textit{Fault Tolerant Elastic Resource Management} (FT-ERM) framework  \cite{saxena2022fault} using multi-resource neural network prediction-based failure estimation. However, prior works are unsuitable for digital twin (DT) tasks due to the intricate correlations among collaborative tasks, which demand augmented reliability, security, and safety. DT applications face several failure modes, including data inconsistencies from delayed or corrupted data, communication failures from network issues, and sensor inaccuracies causing erroneous inputs. Additionally, software bugs, resource contention, and cloud service outages further disrupt DT operations, highlighting the complexities of integrating physical and virtual systems.

To address these challenges, we developed a \textit{SimiFed} prediction unit that facilitates joint learning across interdependent workloads. This approach aggregates similar models to create a federated learning-based global prediction model (FedSim) \cite{palihawadana2022fedsim}, which has been utilized in various fields to enhance security \cite{wu2022coupled,song2022personalized}. For instance,  Awan et al. \cite{awan2023privacy} have utilized FedSim approach for improving privacy-preservation of big data security for internet-of-things (IoT) environments. The proposed method integrates \textit{Cosine Similarity} and \textit{Long Short Term Memory} (\textit{LSTM})-\textit{Federated Learning}, providing a comprehensive solution in the form of \textit{ fault-tolerant and self-healing cloud-based digital twin processing management} (SF-DTM) model. By utilizing a collaborative resource estimation unit with Federated Learning and Cosine Similarity (SimiFed), it effectively addresses data inconsistencies and communication failures, ensuring precise, real-time data processing and synchronization. SF-DTM also mitigates software bugs and resource contention with a self-healing strategy, employing fault-prone pattern analytics to optimize VM allocation. Additionally, it rapidly addresses cloud service outages, minimizing downtime and ensuring continuous operation.

%failure estimation which require precise learning of resource utilization correlations among collaborative tasks of DT applications, when tasks across multiple workload assignments which are interrelated and which demands higher reliability and security and safety from all vulnerabilities during execution. To overcome these challenges the proposed SimiFed prediction unit incorporate advantage of jointly learning across multiple co-interdependent workloads, enabling the global model to leverage most admissible shared information with data security and improving fault prediction accuracy for DT applications. 

%Dai et al. \cite{dai2020deep} have proposed a DT network paradigm for IIoT systems, leveraging Lyapunov optimization and asynchronous actor-critic algorithm for stochastic computation offloading and resource allocation to minimize long-term energy consumption. A  joint optimization and adaptive particle swarm with genetic algorithm-based parallel intelligence-driven resource scheduling scheme is proposed in \cite{yang2023parallel} for computation tasks in intelligent vehicular systems while addressing delay and load balance issues.

\subsection{Paper Contributions}
%In the context of the existing considerable approaches, it is revealed that the fault-tolerant resource allocation and management of cloud empowered collaborative DT development is still at infancy stage and needs a robust solution. To the best of the authors' knowledge,   this is a first paper which proposed a comprehensive solution named as SF-DTM model, for fault tolerant execution and management of DT applications, providing a 360-degree approach towards addressing  the aforementioned critical challenges effectively. Additionally, SF-DTM model preserves the data privacy by ensuring raw data remains local and leverages federated learning with cosine similarity for secure collaborative processing. This approach maintains data integrity and confidentiality, protecting sensitive information during resource sharing within multi-tenant cloud environment. 
Despite existing approaches, fault-tolerant resource allocation and management in cloud-empowered collaborative DT development remain in their infancy and require robust solutions. This paper introduces the SF-DTM model, the first comprehensive solution for fault-tolerant execution and management of DT applications, offering a 360-degree approach to these challenges. Additionally, the SF-DTM model ensures data privacy by keeping raw data local and employs federated learning with cosine similarity for secure collaborative processing, maintaining data integrity and confidentiality in a multi-tenant cloud environment.
Specifically, the key contributions of the paper are threefold:
\begin{itemize}
    \item A novel \textit{SimiFed: Federated Learning with Cosine Similarity Integration} strategy is introduced, incorporating a \textit{collaborative processing estimation} approach that utilizes secure data model learning and sharing. This method enables optimal computation of processing requirements for diverse DT tasks, effectively addressing failures caused by resource congestion stemming from unpredictable and varied real-time resource demands.

    \item A novel \textit{self-healing and fault-tolerant strategy} is proposed which generates frequent sequence knowledge patterns analytics for deciding fault-tolerant DT tasks allocation. The self-healing component is induced by engaging VM replicas-based on Multi-version programming.

    \item Implementation and evaluation of SF-DTM model
using real workload traces reveals its worthiness in executing and managing collaborative  DT applications with high availability and reliability over prior state-of-the-arts.

\end{itemize}

\subsection{Paper Organization}
Section \ref{proposed approach} presents the proposed solution approach  describing SimiFed: DT application processing estimation (Section \ref{s1}); self-Healing and fault-tolerant strategy (Section \ref{s2}); DT tasks assignment and execution (Section \ref{s3}); and operational design and complexity analysis (Section \ref{s4}). The detailed description of the performance evaluation and comparison is discussed in Section \ref{performance evaluations}. Section \ref{conclusion} summarizes the conclusions and outlines future research directions. Table \ref{table:notation} shows the list of symbols with their explanatory terms used throughout the paper.
\begin{table}[htbp]
	\centering
	
	\caption[Table caption text] {Notations and their descriptions}  %\cite[p.10]{refid} }
	\label{table:notation}
%	\resizebox{1.0\textwidth}{!}{\begin{minipage}{\textwidth}
\resizebox{0.35\textwidth}{!}{
			\begin{tabular}{|l|l|}
				\hline
				%\multicolumn{2}{c}{Item} \\
				%\cline{1-2}
							
				$n$& number of clients \\  \hline
				$C$ & client  \\  \hline
 				$A$ & digital twin application  \\  \hline 
			$a$ & component task of DT application  \\  \hline %$LM$ & local model \\  \hline
   %$GM$ & global model \\  \hline 	
   $\theta$ &  global model parameter \\  \hline 
   $D$ & set of data samples  \\  \hline 
   $w_k$ & weight of $k^{th} $ local model\\  \hline 
   $R$ & resource usage  \\  \hline 
   $\eta$ & learning rate  \\  \hline 
   $\Xi$ & mean of absolute error \\  \hline
   $a_j(\Omega)$ & status of $j^{th} $ DT task \\  \hline 
   $R(\Phi)$ & available resource capacity  \\  \hline 
   $R(\Phi^\ast)$ & threshold  resource capacity  \\  \hline 
   $X$, $Y$ & patterns  \\  \hline 
   $\Upsilon$ & mapping between VM and component task ($a$)  \\  \hline 
   $\omega$ & mapping between VMs and servers  \\  \hline 
   $Q$ & number of VMs \\  \hline 
   $P$ & number of servers \\  \hline 
   $f(i)$ & failure probability of $i^{th} $ DT task  \\  \hline 
				
			\end{tabular}}
%	\end{minipage}}
\end{table}

\section{Proposed Model} \label{proposed approach}
\begin{figure}[!htbp]
    \centering
    \includegraphics[width=1.0\linewidth]{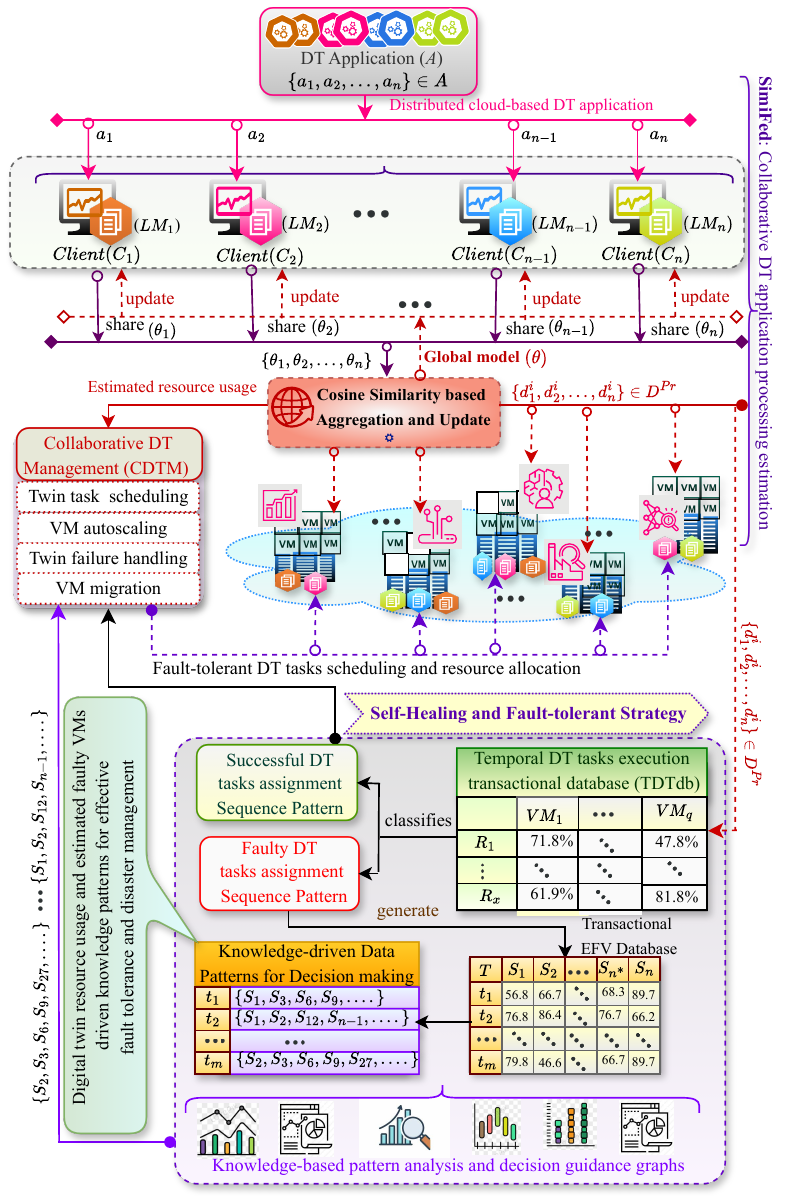}
    \caption{SF-DTM Model}
    \label{fig:proposed}
\end{figure}

Consider a DT application, denoted as $A$, under development and operation by a collaboration of $n$ clients, \{$C_1$, $C_2$, ..., $C_n$\}, spanning various geographical locations, as depicted in Fig. \ref{fig:proposed}. Each of the $n$ components constituting $A$ such that \{$a_1$, $a_2$, ..., $a_n$ $\in A$\}, is executed and managed by its corresponding client operator, \{$C_1$, $C_2$, ..., $C_n$\}, on their respective local machines. The  $n$ components:  \{$a_1$, $a_2$, ..., $a_n$\} are coordinated and comprehensive DT application ($A$) is executed at cloud platform, wherein, it is consistently updated and actively monitored for real time processing. A \textit{Collaborative DT application processing estimation} unit is developed by merging Federated Learning with Cosine Similarity (\textit{SimiFed}). It actively monitors and records resource usage (e.g., processing, storage, networking) at each local site, training respective LSTM-based Local Models \{$LM_1$, $LM_2$, ..., $LM_n$\}. These models are periodically sent to a collaborative Global Model ($GM$) on the cloud for retraining with updated data. During aggregation and updating, similar local models in resource utilization are selected by incorporating \textit{Cosine Similarity} to enhance the global model's accuracy. The global model analyzes estimated resource usage for the collaborative DT application, guiding the \textit{Collaborative DT Management} (CDTM) unit in decision-making about DT task scheduling, fault mitigation, failure handling, resource distribution, and VM migration and autoscaling. Based on the DT application's resource usage, CDTM unit allocates resources for scheduling and execution. Additionally, it employs a Self-Healing and Fault-tolerant Strategy for efficient and reliable real-time execution of task components. The detailed descriptions of SimiFed-based collaborative DT application processing estimation, the self-healing and fault-tolerant strategy, and task scheduling with VM autoscaling, are provided in subsequent Section \ref{s1}, Section \ref{s2}, and Section \ref{s3}, respectively.  

\subsection{SimiFed: DT application processing estimation} \label{s1}

Let  $D$ denote the set of data samples of resource usage of \{$a_1$, $a_2$, ..., $a_n$\} $\in A$, partitioned across \( n \) clients. Each client has a $k^{th}$ subset of the application resource usage data denoted as \( D_k \). Let the \( \theta \) be the global model parameters, \( \theta_k \) represents the  local model parameters at client \( k \), \( f \) is an objective function to be minimized, and \( \nabla f({D_k(t)}_{t=1}^{T}, \theta_k) \) is the gradient of the objective function with respect to the local model parameters at \( k^{th} \) client over \( t = 1 \) to \( T \). The objective of this collaborative learning is to optimize the global model \( \theta \) by aggregating the most admissible local updates from clients while preserving privacy of DT application ($A$). Correspondingly, the federated learning process is formulated as an optimization problem in Eq. (\ref{e1}), which aims to minimize the global objective function \( SimiFed(\theta) \) with respect to the global model parameters \( \theta \). Here, \( n \) represents the number of local models or clients, \( w_k \) are the weights for each local model reflecting its contribution to the global model, and \(\text{LSTM}(\{D_k(t)\}_{t=1}^{T}; \theta_k)\) is the LSTM function approximator for the \( k^{th} \) local model, accounting for the entire sequence of data \(\{D_k(t)\}_{t=1}^{T}\) over time \( t = 1 \) to \( T \).

%Here, the term \(K\): number of local models or clients; \(w_k\): weights for each local model, reflecting its contribution to the global model; \(\text{LSTM}(\{D_k(t)\}_{t=1}^{T}; \theta_k)\): The LSTM function approximator for the \(k^{th}\) local model, taking into account the entire sequence of data \(\{D_k(t)\}_{t=1}^{T}\) over time \(t = 1\) to \(T\).

\begin{equation}
\text{SimiFed}(\theta) = \sum_{k=1}^{n} w_k \cdot \text{LSTM}(\{D_k(t)\}_{t=1}^{T}; \theta_k) \label{e1}  
\end{equation}

The LSTM processes the sequence of data points \(\{D_k(t)\}_{t=1}^{T}\), capturing temporal dependencies, accounting for autocorrelation with its evolving hidden state, and learning long-term trends for predicting dynamic DT patterns in real-time. The cosine similarity function (Eq. \ref{eq:cosineSimilarity}), where, $R$, $R_i$, and $R_j$ represent the resource usage requirements of DT application, $i^{th}$ task, and $j^{th}$ task, respectively, such as \{$R_i, R_j\} \in R$, is employed to group and select the most suitable collaborative task segments. These segments exhibit similarity in their processing requirements, enhancing efficiency and effectiveness in task allocation and resource management.
\begin{eqnarray}
\label{eq:cosineSimilarity}
Cosine(R_i,R_j)& = & \dfrac{R_i \cdot R_j}{||R_i|| ~||R_j||}
\end{eqnarray}
The global model parameters at cloud platform are updated iteratively by aggregating the selected most admissible and similar local updates \{$\widehat{LM_1}$, $\widehat{LM_2}$, ..., $\widehat{LM_{\widehat{n}}}$\} from $\widehat{n}$ clients, where $1 < \widehat{n} \le n $. In each iteration, the central server at cloud platform broadcasts the current global model parameters \( \theta \) to all clients, and each client computes its local update using its local dataset and the received global parameters. The server then aggregates these updated local models to update the global model. The resource estimation involves three consecutive steps: \textit{Firstly}, the global model parameters \( \theta \) are initialized. \textit{Secondly}, each client \( i \) computes its local update \( \Delta\theta_i \) using the received global parameters as stated in Eq. (\ref{e2}),  where \( \eta \) is the learning rate. The central server aggregates the selected local updates to obtain the new global parameters using Eq. (\ref{e3}). \textit{Thirdly}, the second step of communication is repeated until convergence criteria are met. This process learning  preserves the privacy of DT application execution as the engaged data remains decentralized and never leaves the clients' devices.
    \begin{gather}
       \Delta\theta_i = -\eta \nabla f(\text{LSTM}({D_i(t)}_{t=1}^{T}, \theta_i ) \label{e2} \\
        \theta \leftarrow \theta + \sum_{k=1}^{\widehat{n}} \frac{|D_k|}{|D|} \cdot \Delta\theta_k \label{e3}
    \end{gather}
The outcome of the resource estimation model is analyzed and evaluated using the mean ($\Xi$) of absolute error ($AE$) and mean squared error score  (${MSE}$) using Eq. (\ref{mean}) and Eq. (\ref{rmse}), respectively for $m$  resource usage samples, wherein ${D}_i^{Ac}$ and ${D}_i^{Pr}$ are respective actual and predicted resource usage requirement of $i^{th}$ DT task.  
%%%%%%%%%%%%%%%%%%%%%%%%%%%%%%%%%%%%%
%\begin{equation}
\begin{gather}
	\label{mean}
	\bar{\Xi} = \frac{\sum_{i=1}^{m}AE}{m} \\ 
	\label{rmse}
{MSE} = {\frac{1}{m}\sum_{i=1}^{m}({D}_i^{Ac}-{D}_i^{Pr})^2} 
%\end{equation}
\end{gather}    
   
\subsection{Self-Healing and Fault-tolerant strategy} \label{s2}
As illustrated in Fig. \ref{fig:proposed}, \textit{Self-Healing and Fault-tolerant strategy} unit receives estimated  usage of $i^{th}$ resource: ($D^{Pr}$) from SimiFed estimator such that \{$d^i_1$, $d^i_2$, ..., $d^i_n$ $\in D^{Pr}$\}, wherein, $d^i_j$ represents estimated usage of $i^{th}$ resource of $j^{th}$ DT application ($A$). Let  the DT task components \{$a_1$, $a_2$, ..., $a_n$ $\in A$\} are executed on $n$ VMs hosted on $P$ different servers.  A relational temporal database named `\textit{Temporal DT tasks execution
 transactional database}' (TDTdb) is prepared for examination and comparison with the available $i^{th}$ resource capacity ($R^i(\Phi)$) of VM to compute failure probability of different DT task components \{$a_1$, $a_2$, ..., $a_n$\}. It utilizes Eq. (\ref{eq:case}) to determine status ($a_j(\Omega)$) of $j^{th}$ DT task component as \textit{Highly fault-prone} ($a^{\ast}_j$), \textit{Mild fault-prone} (${\bar{a}}_j$), and \textit{Least fault-prone} ($a^{\dagger}_j$) based on the comparison of estimated resource usage ($d^i_j$) with threshold ($R^i_j(\Phi^\ast)$) and available ($R^i_j(\Phi)$) resource capacity of $i^{th}$ resource of $j^{th}$ VM. Accordingly, the highly fault-prone ($a^{\ast}_j$) and mild fault-prone (${\bar{a}}_j$) tasks are considered to be \textit{fault-prone DT task components} represented as \{$a^{\ast}_1$, $a^{\ast}_2$, ..., $a^{\ast}_{n^{\ast}}$\} while least fault-prone ($a^{\dagger}_j$) tasks are identified as \textit{efficient DT task components} \{$a^{\dagger}_1$, $a^{\dagger}_2$, ..., $a^{\dagger}_{n^{\dagger}}$\} such that \{$a^{\ast}_1$, $a^{\ast}_2$, ..., $a^{\ast}_{n^{\ast}}$ $\cup$ $a^{\dagger}_1$, $a^{\dagger}_2$, ..., $a^{\dagger}_{n^{\dagger}}$\} $\in A$. 
\begin{equation} \label{eq:case}
        a_j(\Omega) = \begin{cases}
                   \textit{Highly fault-prone} (a^{\ast}_j)   &  {if(d^i_j > R^i_j(\Phi^\ast) < R^i_j(\Phi))} \\
                    \textit{Mild fault-prone}({\bar{a}}_j)& {if(d^i_j = R^i_j(\Phi^\ast) < R^i_j(\Phi))} \\
                \textit{Least fault-prone}(a^{\dagger}_j) & \textit{Otherwise }
                    \end{cases}
\end{equation}

\begin{figure} [!htbp]
    \centering
    \includegraphics[width=1.0\linewidth]{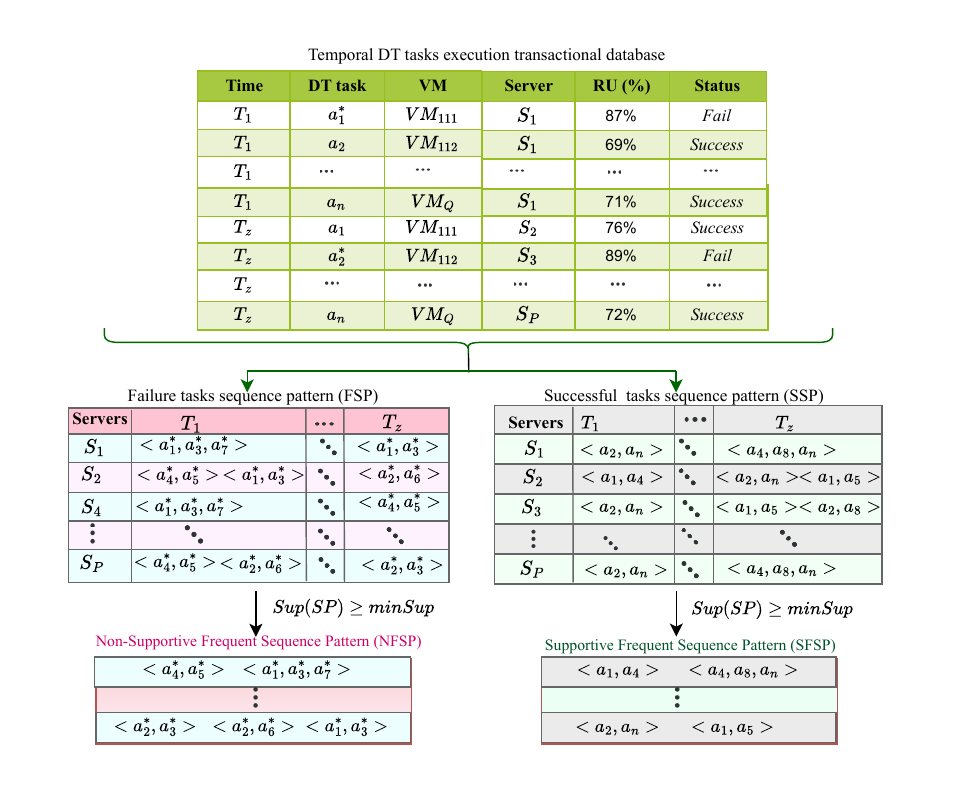}
    \caption{Frequent sequence pattern analytics}
    \label{fig:pattern}
\end{figure}

TDTdb reports list of DT tasks: \{$a_1$, $a_2$, ..., $a_n$\}, VMs used for respective tasks execution \{$VM_{111}$, $VM_{112}$, ..., $VM_{Q}$\}, servers  \{$S_1$, $S_2$, ..., $S_P$\}, and respective tasks resource usage over consecutive time instances \{$T_1$, $T_2$, ..., $T_z$\} as demonstrated in Fig. \ref{fig:pattern}. TDTdb contains a union set of failure  and successful DT tasks as stated in Eq. (\ref{p1}), which comprises  different  failed DT task components: \{$a^{\ast}_1$, $a^{\ast}_2$, ..., $a^{\ast}_{n^{\ast}}$\} and successful tasks: \{$a_1$, $a_2$, ..., $a_n$\}, respectively observed over consecutively ordered set of $z$ timestamps \{$T_1$, $T_2$, ..., $T_{z}$\}. 
TDTdb is analysed to produce \textit{Faulty DT tasks assignment Sequence Pattern} ($FSP$) knowledge database and \textit{Successful DT tasks assignment Sequence Pattern} ($SSP$) knowledge database, respectively. They contain $K^\ast$ sequence patterns ($SP$) of $n^{\ast}$ failed DT task components: \{$\langle a^{\ast}_{it1}, a^{\ast}_{it2}, \ldots, a^{\ast}_{itK^{\ast}}\rangle  $\} (Eq. (\ref{p2})) and $K$ sequence patterns ($SP$) of  $n$ successful tasks: \{$\langle a_{it1}, a_{it2}, \ldots, a_{itK}\rangle $\} (Eq. (\ref{p3})), respectively,  on $Q$ server machines \{$S_1$, $S_2$, ..., $S_Q$\} observed during $z$ timestamps. 
\begin{gather}
TDT =\bigcup_{t=T_1}^{T_z} (\{a_1, a_2, ..., a_n\}(t) \cup \{a^{\ast}_1, a^{\ast}_2, ..., a^{\ast}_{n^{\ast}}\}(t))  \label{p1} \\
FSP (t) = \bigcup_{i=1}^{n^{\ast}}  \bigcup_{t=T_1}^{T_2} \{ \langle a^{\ast}_{it1}, a^{\ast}_{it2}, \ldots, a^{\ast}_{itK^{\ast}}\rangle \}  \label{p2}  \\
%\{\langle a^{\ast}_i, a^{\ast}_2, ..., a^{\ast}_{n^{\ast}} \rangle \} (t) \quad t \in \{T_1, T_2, ..., T_z\}\label{p2}  \\
SSP (t) = \bigcup_{i=1}^{n}  \bigcup_{t=T_1}^{T_2} \{ \langle a_{it1}, a_{it2}, \ldots, a_{itK}\rangle  \}  \label{p3} 
\end{gather}

Let a sequence pattern ($SP$) of co-assignment of faulty DT task and successful DT task components  on a common server machine are represented as $\langle$\{$a^{\ast}_1$, $a^{\ast}_3$, $a^{\ast}_7$\}, \{$a^{\ast}_4$, $a^{\ast}_5$\}, ..., \{$a^{\ast}_2$, $a^{\ast}_3$\}$\rangle$  and $\langle$\{$a_2$, $a_n$\}, ...,\{$a_4$, $a_8$, $a_n$\}$\rangle$, respectively, as illustrated in Fig. \ref{fig:pattern}.
 A  sequence pattern ($SP$) is defined as a \textit{Periodic Frequent Sequence Pattern} (PFSP)  when it repeats with a significant frequency after a periodic time duration,   such that $Sup(SP) \ge minSup$, where $Sup(SP)$ is support or frequency of occurrence of $SP$ and $minSup$ refers to the minimum support value. Accordingly, two types of PFSP ($P^{St}$) are determined to generate more relevant knowledge pattern analytics including \textit{Non-supportive Frequent Sequence Pattern} ($Nf$) based on the analysis of failed DT tasks assignment sequence patterns and \textit{Supportive Frequent Sequence Pattern}  ($Sf$) driven from Successful DT tasks assignment sequence patterns using Eq. (\ref{p5}). The term $|FSP\bigcap \{X_i\}|$ denotes the cardinality or count of the intersection of set $FSP$ and the set containing only specific $i^{th}$ pattern $X_i$ while $|SSP\bigcap \{Y_j\}|$ refers the count of the intersection of set $SSP$ and the set containing only specific $j^{th}$ pattern $Y_j$. These patterns are distinguished on the basis of the correlation observed between resource usage of various DT tasks and their assignment on specific server machines having varying resource capacity. %Henceforth, the significant knowledge achieved from NFSP and SFSP analysis is utilized to minimize the probable faults or resource congestion-based failures of fault-prone DT tasks \{$a^{\ast}_1$, $a^{\ast}_2$, ..., $a^{\ast}_{n^{\ast}}$\} by maximizing SFSP driven tasks allocation and minimizing NFSP guided tasks assignment.

\begin{equation}\label{p5}
    \begin{gathered}
     P^{St}  = \begin{cases}
                   \{Nf: Nf \bigcup X_i\}  &  {if(|FSP\bigcap \{X_i\}| \ge minSup_i} \\
                \{Sf: Sf \bigcup Y_j\}  &  {if(|SSP\bigcap \{Y_j\}| \ge minSup_j} \\  
                \textit{Insignificant} & \textit{Otherwise }
                    \end{cases}
                    \\
                \forall_i: i \in [1, K^\ast],  \forall_j: j \in [1, K]   
    \end{gathered}
\end{equation}

\subsection{Assignment and Execution} \label{s3}
The efficient DT tasks \{$a^{\dagger}_1$, $a^{\dagger}_2$, ..., $a^{\dagger}_{n^{\dagger}}$\} are executed on selected suitable VMs which satisfies their respective resource demands based on the First-Fit Decreasing policy of tasks assignment. The essential constraints that must be satisfied for DT tasks assignment and execution are stated in Eqs. (\ref{task} and \ref{VMP}). The term $\Upsilon_{ki}$ is a mapping $\Upsilon_{ki}: a^{\dagger}_k \times VM_i \in \{1, 0\}$ such that $\Upsilon_{ki}= 1$, if $k^{th}$ task $a^{\dagger}_k$ is assigned to VM $VM_i$, else, it is $0$;  $\forall i \in [1, Q], k \in [1, n^{\dagger}]$. Similarly, $\omega_{li}$ represents a mapping $\omega_{li}: VM_i \times S_l \in \{1, 0\}$ such that $\omega_{li}= 1$, if VM $VM_i$ is deployed on server $S_l$, else, it is $0$;  $\forall i \in [1, Q], l \in [1, P]$;  ${{R_j}}$ specifies $j^{th}$ resources viz., CPU and memory for assignment of VM ($VM_i$) on $S_l$. 
\begin{gather}
\sum_{k=1}^{n^{\dagger}}{a^{\dagger}_k\times {{{R}_j}} \times \Upsilon_{ki} \le VM_i\times {{{R}_j}}} \label{task} \\
\sum_{i=1}^{Q}{VM_i\times {{{R}_j}} \times \omega_{li} \le S_l\times {{{R}_j}}}  \label{VMP}
\end{gather}

To accomplish fault-tolerant DT tasks assignment, an odd number \{$2x + 1: x \ge 1$\} of replicas of fault-prone DT tasks \{$a^\ast_1$, $a^\ast_2$, ..., $a^\ast_n$\} are assigned and executed concurrently subject to execution cost and service agreement constraints. Let multiple version programming (MVP)-based fault-tolerance is employed with an odd number of active images or versions of DT task $a_i$ simultaneously.  The failure probability of MVP ($\mathds{F}^{MVP}$), can be computed using Eq. (\ref{eq:mvp}), where \( num \) represents the number of versions, and \( f(i) \) represents the failure probability of alternative DT task $a_i$. 

\begin{equation} \label{eq:mvp}
\mathds{F}^{MVP} = \sum_{i=\frac{num+1}{2}}^{num} f(i)
\end{equation}

The expression \( \frac{num+1}{2} \) denotes the midpoint of the range of versions, ensuring that MVP fails if and only if the number of failed VM images exceeds the majority threshold. In this process, the valuable insights gained from NFSP and SFSP analysis are leveraged to reduce the likelihood of faults or failures due to resource congestion in fault-prone DT tasks {$a^{\ast}_1$, $a^{\ast}_2$, ..., $a^{\ast}_{n^{\ast}}$} by maximizing the allocation of SFSP-driven tasks and minimizing the assignment of NFSP-guided tasks.
\subsection{Operational Design and Complexity} \label{s4}
Algorithm \ref{algo:alg1} outlines operational flow of SF-DTM. 
\begin{algorithm}[!htbp]
	\caption{SF-DTM Operational summary}
	\label{algo:alg1}
	\textbf{Input} Temporal DT Application database (TDTdb)\;
	\For {each DT task \{$a_1$, $a_2$, ..., $a_n$\} $\in A$}{ 
 Train \{$LM_1$, $LM_2$, ..., $LM_n$\} at $n$ client sites \;
 \For {each communication round \{$i_1$, $i_2$, ..., $i_z$,\}}{
 Share, aggregate and update the global model parameters ($\theta$) using Eq. (\ref{eq:cosineSimilarity}) \;
 Analyse and update global model based on computation of Eqs. (\ref{e2}) and (\ref{e3}) \;
 Estimate collaborative processing requirement: \{$d^i_1$, $d^i_2$, ..., $d^i_n$ $\in D^{Pr}$\} \;
 Examine status of \{$a_1$, ..., $a_n$\} using Eq. (\ref{eq:case}) \;
 }
 Using TDTdb, prepare $FSP$ and $SSP$ using Eq. (\ref{p2}) and Eq. (\ref{p3}) \;
 Apply Eq. (\ref{p5}) to generate $Nf$
 and $Sf$ decision making significant sequence pattern  \;
 Assign DT tasks \{$a^{\dagger}_1$, $a^{\dagger}_2$, ..., $a^{\dagger}_{n^{\dagger}}$\} on VMs subject to constraint stated in Eqs. (\ref{task} and \ref{VMP}) \;
 Create an odd number of replicas of fault-prone DT tasks  and assign them on VMs subject to constraint stated in Eqs. (\ref{task} and \ref{VMP}) while minimizing non-supporting ($Nf$) assignment\;
 }
 \end{algorithm}
Step 1 retrieves TDTdb input, while Steps 2-12 iterate for $n$ DT task executions. Step 3 generates $n$ local models. Steps 4-8 employ the SimiFed unit to predict processing requirements, with time complexity $\mathcal{O}(nT+zU)$ ($n$: number of local models, $T$: LSTM  time complexity, $U$: complexity of updating the global model). Step 9 updates the database in $\mathcal{O}(1)$ time. Steps 10 and 11 use the frequent sequence pattern mining algorithm with $\mathcal{O}(D^S)$ complexity ($D$: number of distinct items, $S$: length of the longest sequence). Finally, Step 12 has a complexity of $\mathcal{O}(1)$. Overall computational complexity: $\mathcal{O}(nT+zU + D^S)$.

\section{Performance Evaluation} \label{performance evaluations}
\subsection{Experimental set-up and Dataset}
The simulation experiments are executed on a server machine assembled with two Intel\textsuperscript{\textregistered} Xeon\textsuperscript{\textregistered} Silver 4114 CPU with 40 core processor and 2.20 GHz clock speed. The server machine is deployed with 64-bit Ubuntu 16.04 LTS, having main memory of 128 GB. The data centre environment is implemented in Python including three types of servers and four types of VMs configuration shown in Tables \ref{table:server} and \ref{table:vm}. %The resource features like power consumption ($PW_{max}, PW_{min}$), MIPS, RAM and memory are taken from real server IBM \cite{IBM1999} configuration where $S_1$ is `ProLiantM110G5XEON3075', $S_2$ is `IBMX3250Xeonx3480' and $S_3$ is `IBM3550Xeonx5675'. The VMs configuration is inspired from the VM instances of Amazon website \cite{amazon1999EC2}. 
 
 \begin{table}[!htbp]
 	\centering
 	
 	\caption[Table caption text] {Server Configuration}  %\cite[p.10]{refid} }
 	\label{table:server}
 	%\resizebox{0.8\textwidth}{!}{\begin{minipage}{\textwidth}
 	\resizebox{8.5cm}{!}{
 		\begin{tabular}{lccccc}
 			\hline
 			%\multicolumn{2}{c}{Item} \\
 			%\cline{1-2}
 			Server&PE&MIPS&RAM(GB)&$\mathds{PW}_{max}$&$\mathds{PW}_{min}$/$\mathds{PW}_{idle}$\\
 			\hline
 			$S_1$ 	& 2&2660&4&135&93.7 \\
 			$S_2$	& 4&3067&8&113&42.3 \\
 			$S_3$	& 12&3067&16&222&58.4 \\

 			\hline
 	\end{tabular}}
 	%	\end{minipage}}
 \end{table}
 
 \begin{table}[!htbp]
 	\centering
 	
 	\caption[Table caption text] {VM Configuration}  %\cite[p.10]{refid} }
 	\label{table:vm}
 	%	\resizebox{0.8\textwidth}{!}{\begin{minipage}{\textwidth}
 	\begin{tabular}{lccc}
 		\hline
 		%\multicolumn{2}{c}{Item} \\
 		%\cline{1-2}
 		VM type& PE &MIPS&RAM(GB)\\
 		\hline
 		$VM_{small}$&1&500&0.5\\
 		$VM_{medium}$&2&1000&1\\
 		$VM_{large}$&3&1500&2\\
 		$VM_{Xlarge}$&4&2000&3\\

 		\hline
 	\end{tabular}
 	%	\end{minipage}}
 \end{table}

To evaluate the SF-DTM model, we needed the resource requirements of different component tasks in varying sizes of DT applications. This information was crucial to train the proposed SimiFed collaborative workload estimation unit and to develop the DT fault transactional database (TDTdb). According to the requirement of DT manufacturing and processing applications, which involve remote collaboration on diverse tasks at different sites, it was imperative to train the SimiFed unit to learn the relative resource requirements of different processing tasks within a DT application. To achieve this, we used VMs data traces of benchmark Google Cluster Workload (GCW)  \cite{reiss2011google}, which provided CPU, memory, and disk I/O usage information for 672,300 jobs executed on 12,500 servers over a 29-day period. The CPU and memory utilization of the virtual machines were derived from the observed usage percentages every five minutes over a 24-hour period. \par
For each experimental case, local training models based on LSTM were used, using ReLU activation, Adam optimizer, and a final softmax output layer. Training and testing were conducted using an 80:20 split data set for analysis, periodically in real time.  The experiments were carried out on multiple scenarios involving DT applications with collaborative task counts ranging from 10 to 100 in increments of 10. Accuracy, loss values, and calibrations were recorded after each global model optimization. These tasks were scheduled to be executed on various VMs hosted on available physical machines, randomly at runtime. Collectively, these tasks formed a DT application, with each task assumed to be operated from remote sites where local models, corresponding to the resource requirements of different tasks, were trained separately. Subsequently, Cosine Similarity was employed to identify the most appropriate and similar resource requirement local models and consequently build global model. In the subsequent phase of the same experimental cases, TDTdb was utilized to create fault patterns for each DT application exclusively. These fault patterns were then used to train the fault frequent sequence pattern mining and analysis unit with various minimum support values. 
\subsection{Key Performance Indicators}
The performance of SF-DTM is evaluated in terms of 
MTBF (Eq. \ref{mtbf}), MTTR (Eq. \ref{mttr}), availability ($\mathds{AV}$ in Eq. (\ref{availability}),  resource utilization ($\mathds{RU}$), and power consumption ($\mathds{PW}$) using Eqs. (\ref{RU1}) and (\ref{RU2}), and Eq. (\ref{power2}), respectively. 
\begin{gather}
\label{mtbf}
MTBF=\int\limits_{\substack{t_1\\\mathcal{}}}^{t_2}(\frac{\sum_{i=1}^{M}{UT_i}}{Num_{\mathds{F}}})dt
\\
\label{mttr}
MTTR= \int\limits_{\substack{t_1\\\mathcal{}}}^{t_2}(\frac{\sum_{i=1}^{M}{DT_i}}{Num_{\mathds{F}}})dt
\\
\label{availability}
\mathds{AV}_{avg} = \frac{MTBF}{MTBF+MTTR}
\\
\mathds{RU}= \frac{\sum_{k=1}^{P}{\mathds{RU}_k^{C}} + \sum_{k=1}^{P}{\mathds{RU}_k^{Mem}}}{|\mathds{N}| \times \sum_{k=1}^{P}{\beta_k} } \label{RU1} \\
\mathds{RU}_k^{\mathds{R}} = \frac{\sum_{i=1}^{Q}{\omega_{ik}} \times V_i^{\mathds{R}}}{S_k^{\mathds{R}}} \quad \forall_k \in \{1, P\} \label{RU2}	\\
\mathds{PW} = 
 \sum_{i=1}^{P} {[{\mathds{PW}_i}^{max} - {\mathds{PW}_i}^{min}]\times{RU} + {\mathds{PW}_i}^{idle}}
 \label{power2}
\end{gather}
In Eqs. (\ref{mtbf}-\ref{availability}), the term $Num_{F}$ represents the total number of failures, while $\sum_{i=1}^{M}{UT_i}$ and $\sum_{i=1}^{M}{DT_i}$ denote the total uptime and downtime experienced by $M$ clients during the time interval [$t_1$, $t_2$]. In Eqs. (\ref{RU1} and \ref{RU2}), $\mathds{N}$ stands for the number of resources, with $\mathds{RU}^{C}$ and $\mathds{RU}^{Mem}$ representing the CPU and memory of a server, respectively. If the $k^{th}$ server $PM_k$ is active (hosting VMs), $\beta_k$ equals 1; otherwise, it equals 0. In Eq. (\ref{power2}), ${\mathds{PW}_i}^{max}$, ${\mathds{PW}_i}^{min}$, and ${\mathds{PW}_i}^{idle}$ denote the maximum, minimum, and idle state power consumption, respectively, of the $i^{th}$ server.

\subsection{Results and Discussion}

Table \ref{table:performanceGCD} presents performance metrics for varying sizes of DT applications over a 400-minute period. These metrics include MTBF, MTTR, availability ($\mathds{AV}$), fault prediction accuracy ($\mathds{F}^{Pred}$), resource contention ($\mathds{RC}$ \%), VM migration ($\mathds{MIG}$ \#), power consumption ($\mathds{PW}$), resource utilization ($\mathds{RU}$), number of overloads ($\mathds{OV}$\%), and load allocation success rate ($\mathds{SUC}$\%). 
  \begin{table*}[!htbp]
 	\caption[Table caption text] {Performance metrics for fault-tolerant DT collaborative application execution} 
  \centering
 	\label{table:performanceGCD}
 	\small
 	%\centering 
 	%\resizebox{0.8\textwidth}{!}{\begin{minipage}{\textwidth}
 	\resizebox{14cm}{!}{
 		
 		\begin{tabular}{|l|c|c|c|c|c|c|c|c|c|c|c|}
 			\hline
 			
 		\textit{Size}($A$) &$T(min.)$& $MTBF$&$MTTR$& $\mathds{AV}$(\%)&$\mathds{F}^{Pred}$(\%) &$\mathds{RC}$(\%) & $\mathds{MIG}$ \#& $\mathds{PW}$ (KW) & $\mathds{RU}$ (\%) & $\mathds{OV}$(\%)& $\mathds{SUC}$(\%) \\ \hline \hline			
 			\multirow{3}{*}{10}
 	     	&50& 857.34& 0.041&99.66& 92.68&7.41&42& 1.554&73.43&2.4 &97.6 \\ \cline{2-12}
 			&100&912.67 & 0.010 & 99.41& 94.52& 5.44&38&1.554 &73.43 &2.8&97.2 \\ \cline{2-12}
 			&200&1171.83&0.005&99.51&96.63& 3.39&26&1.554 &73.43 &4.2 &95.8 \\ \cline{2-12}
 			&400& 1423.27& 0.003&99.70&92.48& 7.37&31&1.554 &73.43 &3.4&96.6\\ \hline \hline

    \multirow{3}{*}{20}
 	     	&50& 873.89& 0.023&99.14&91.12&8.89& 38&2.45&69.34 &4.6&95.4 \\ \cline{2-12}
 			&100&805.77 & 0.015& 99.82& 92.34& 7.63&41&2.45&69.69  &1.7 &98.3\\ \cline{2-12}
 			&200&1043.45&0.022&98.76&94.60&3.38 &46&2.45&69.69&3.2 &96.8  \\ \cline{2-12}
 			&400& 1211.44& 0.012&99.72&95.24& 4.76&37&2.45&69.69 &4.9&95.1 \\ \hline \hline
    
 	  \multirow{3}{*}{40}
 	     	&50& 869.47& 0.022&99.31& 94.56&5.44 & 61&2.45&70.81&4.6 &95.4 \\ \cline{2-12}
 			&100&773.23 & 0.024 & 98.20& 95.63& 4.37&55&2.45&70.81&2.7 &97.3 \\ \cline{2-12}
 			&200&1027.87&0.190&99.06&93.73&6.26 &47&2.45&70.81&6.3&93.7   \\ \cline{2-12}
 			&400& 1167.79& 0.041&99.31&96.62&3.37&55&2.45&70.81&2.9&97.1 \\ \hline \hline		

      \multirow{3}{*}{60}
 	     	&50& 776.89& 0.021&99.19&92.77&7.23& 51&3.14&70.77&5.6&94.4  \\ \cline{2-12}
 			&100&1027.24 & 0.032 & 99.21& 94.51& 5.50&73&3.14 &70.77&3.7&96.3  \\ \cline{2-12}
 			&200&973.57&0.100&99.34& 95.86& 4.14&65&3.14&70.77&5.2&94.8   \\ \cline{2-12}
 			&400& 1303.67& 0.014&99.16&96.81& 3.19&48&3.14 &70.77&3.6&96.4 \\ \hline \hline	
 	\multirow{3}{*}{80}
 	     	&50& 872.66& 0.024&99.19&97.42&2.58& 86&3.94&69.07&5.6&94.4  \\ \cline{2-12}
 			&100&915.83 & 0.010 & 99.21& 96.34& 3.66&91&3.94 &69.07&3.7 &96.3 \\ \cline{2-12}
 			&200&1048.34&0.025&99.34&94.63& 5.37&79&3.94 &69.08&5.2&94.8   \\ \cline{2-12}
 			&400& 1209.63& 0.033&99.16&94.79& 5.21&62&3.94&69.07&3.6&96.4 \\ \hline	\hline	
    \multirow{3}{*}{100}
 	     	&50& 853.45& 0.023&98.19&93.33&6.67& 104&5.28 &66.98&5.6&94.4  \\ \cline{2-12}
 			&100&869.91 & 0.032 & 99.21& 96.22& 3.78&98&5.28 &66.98&3.7 &96.3 \\ \cline{2-12}
 			&200&1061.98&0.024&99.34&91.55& 8.45&89&5.28 &66.98&5.2&94.8   \\ \cline{2-12}
 			&400& 1208.67& 0.019&99.16&95.05& 4.95&99&5.28 &66.98&3.6&96.4 \\ \hline
 	\end{tabular}}
 \end{table*}
 The MTTR value for a VM, taken from \cite{santos2017analyzing}, is 0.21 minutes. There is an inverse relationship between MTTR and MTBF, where increasing MTBF leads to decreasing MTTR. Consequently, MTTR values are computed for varying numbers of VM migrations ($\mathds{MIG}$ \#), influenced by unforeseen faults or resource contention levels ($\mathds{RC}$ \%). Availability is determined using Eq. (\ref{availability}) over a 400-minute time interval, consistently exceeding 99\% for all observed DT application sizes. The SF-DTM model demonstrates resilience and scalability, maintaining consistent performance despite dynamic shifts in fault prediction accuracy ($\mathds{F}^{Pred}$) and resource contention ($\mathds{RC}$ \%). Notably, system performance remains unaffected by execution time variations. Moreover, observed power consumption ($\mathds{PW}$) and resource utilization ($\mathds{RU}$) values are acceptable, increasing with DT application size but independent of other factors.

\subsubsection{DT processing requirement estimation metrics}
\begin{figure*} [!htbp]
    \centering
    \includegraphics[width=0.7\linewidth]{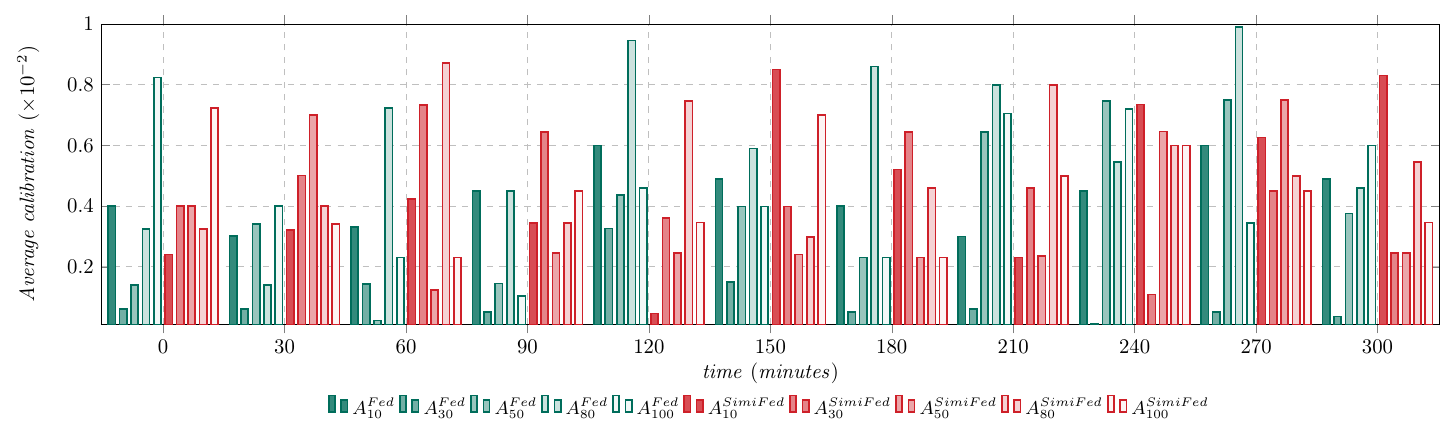}
    \caption{Observed calibration over consecutive time duration of 300 minutes}
    \label{fig:calibration}
\end{figure*}

\begin{figure*}[!htbp]
\centering	
\subfigure[Accuracy v/s epochs]{
\includegraphics[width=0.28\linewidth, scale=8]{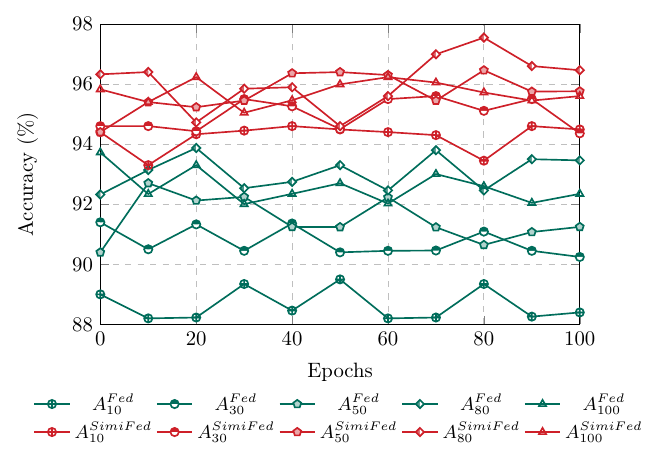}  
}
\subfigure[Loss values v/s epochs ]{\includegraphics[width=0.28\linewidth, scale=8]{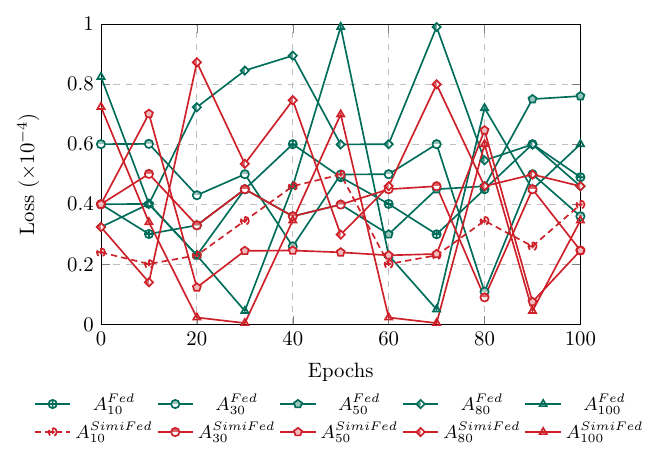}
}
\subfigure[Resource contention prediction]{
  \includegraphics[width=0.28\linewidth, scale=8]{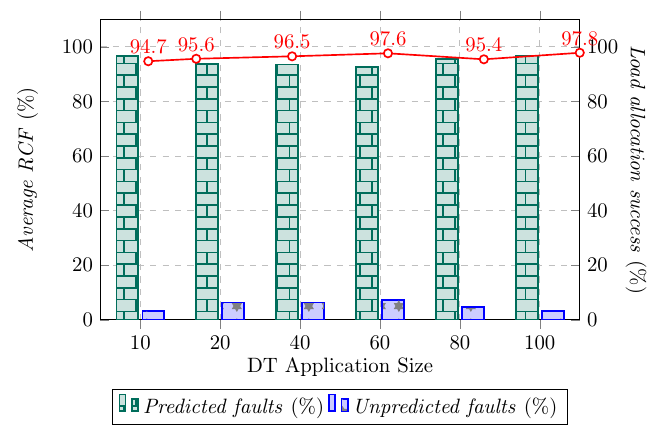}
}
\caption{DT application fault estimation metrics}
%\caption{Resource usage estimation errors}
\label{fig:fault metrics}
\end{figure*}

The processing requirements of DT applications are estimated using the proposed SimiFed prediction unit and it is compared with Federated learning (Fed)-based prediction unit by conducting an extensive range of experiments with varying size such as \{10, 20, ..., 100\} of DT applications. The calibration observed during building of various DT applications models is presented in Fig. \ref{fig:calibration}, which fluctuates slightly within the range of [0.0001 - 0.001] over consecutive 300 minutes.
Fig. \ref{fig:fault metrics} shows the comprehensive performance of resources (viz., CPU, memory etc.) usage forecasting in terms of achieved prediction \textit{accuracy} and \textit{loss values} over 100 consecutive epochs  in Fig. \ref{fig:fault metrics}(a) and Fig. \ref{fig:fault metrics}(b),  respectively, for different DT collaborative applications. Notably, the accuracy score achieved by SimiFed surpasses that of the Fed-based forecasting unit. Additionally, the corresponding loss values exhibit dynamic fluctuations, although consistently remaining marginally lower than those attained through Fed-based learning and forecasting units. The main reason behind this enhanced performance lies in the incorporation of \textit{Cosine Similarity} prior to the construction of the global forecasting model. This strategic inclusion facilitates the selective consideration of the most relevant and similar local models, a capability typically absent in conventional Federated learning (Fed-based) method. As a result, a more precise and accurate DT processing estimation model is constructed, yielding significant improvements in forecasting accuracy. In Fig. \ref{fig:fault metrics}(c), the corresponding prediction values pertaining to resource contention during the allocation and execution phases of DT applications of varying sizes are reported. It is observed that  the load allocation success rate consistently exceeds 94\% across all DT application sizes. However, it is noteworthy that the predicted faults, as indicated by the average of resource contention failure (RCF \%), consistently exceeds 95\%. Conversely, unpredicted faults remain below 4.8\% for all cases.

\begin{figure*}[!htbp]
\centering	
\subfigure[DT Application Size 10]{
\includegraphics[width=0.28\linewidth, scale=8]{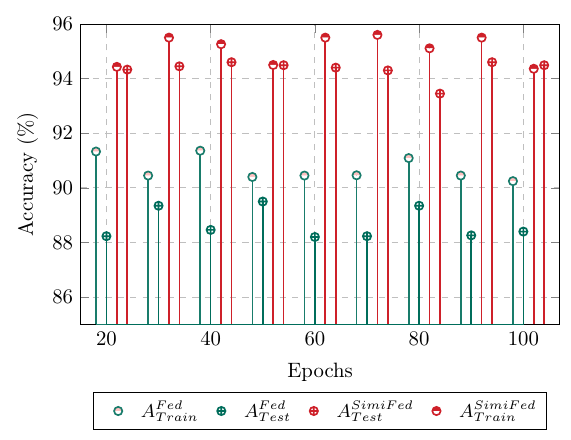}  
}
\subfigure[DT Application Size 50 ]{\includegraphics[width=0.28\linewidth, scale=8]{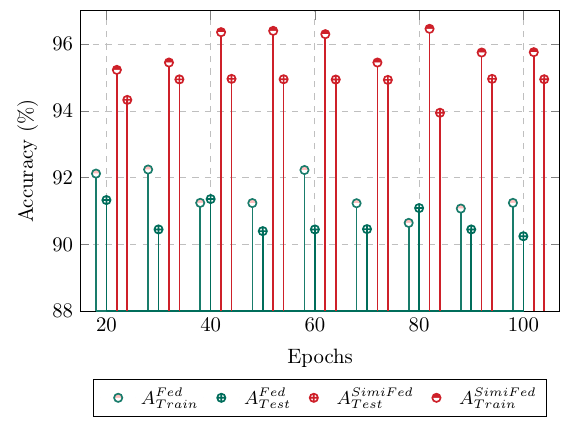}
}
\subfigure[DT Application Size 100]{
  \includegraphics[width=0.28\linewidth, scale=8]{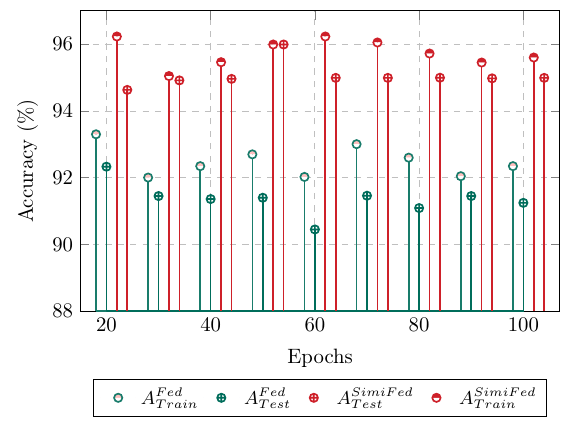}
}
\caption{Fault estimation training accuracy versus testing accuracy for various sizes of DT applications}
%\caption{Resource usage estimation errors}
\label{fig:TT}
\end{figure*}

Fig. \ref{fig:TT} shows training and testing accuracies over 100 epochs for different sizes of DT applications (10, 50, and 100), revealing that training and testing accuracies are notably closer for the SimiFed method compared to the Federated Learning (Fed-based) method.

\subsubsection{Fault pattern generation metrics}
\begin{figure*}[!htbp]
\centering	
\subfigure[Number of Patterns over varying $minSup$]{
\includegraphics[width=0.28\linewidth, scale=8]{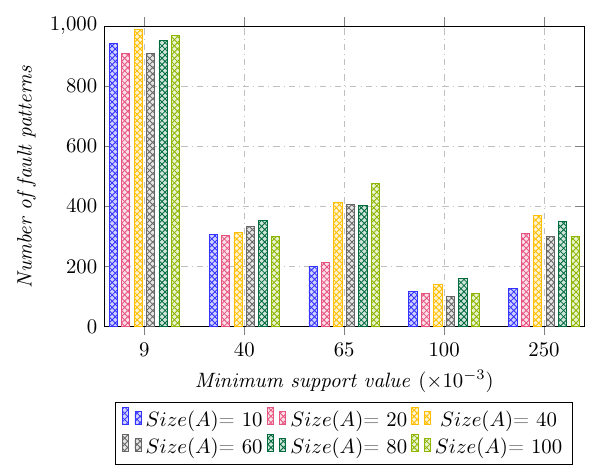}  
}
\subfigure[Runtime consumption]{
  \includegraphics[width=0.28\linewidth, scale=8]{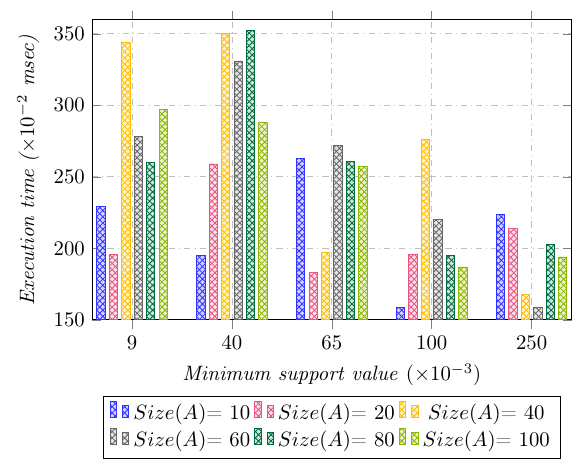}
  }
  \subfigure[Memory space consumption]{\includegraphics[width=0.28\linewidth, scale=8]{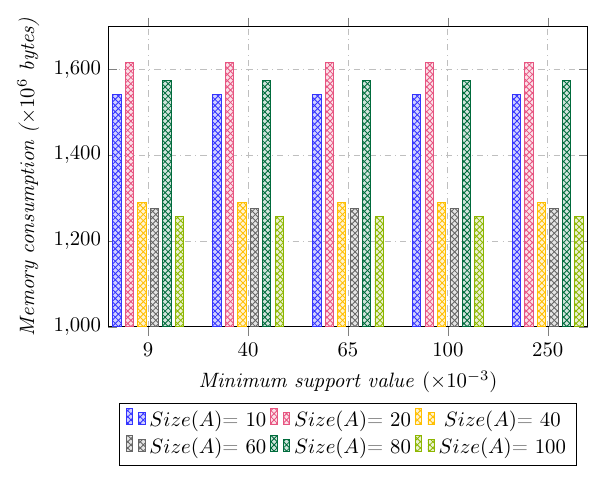}
}
\caption{Fault pattern generation metrics}
%\caption{Resource usage estimation errors}
\label{fig:pattern metrics}
\end{figure*}

In Fig. \ref{fig:pattern metrics}, the average values of performance metrics, including the \textit{number of significant patterns}, \textit{runtime}, and \textit{memory space}, are presented for the generation of frequent sequence-based fault estimation patterns. Fig. \ref{fig:pattern metrics}(a) illustrates the variation in the number of fault patterns obtained across different minimum support ($minSup$) values (0.009, 0.040, 0.065, 0.100, 0.250) for varying sizes of DT applications ($Size(A)$). Notably, the highest number of significant patterns was observed for $minSup$ = 0.009 with $Size(A)$ = 40. The elapsed time and memory usage during the generation of useful fault estimation patterns are depicted in Fig. \ref{fig:pattern metrics}(b) and Fig. \ref{fig:pattern metrics}(c), respectively. It is noteworthy that there is a slight variation in execution time observed with different $minSup$ values across various sizes of DT applications during pattern generation. Conversely, memory consumption remains largely unaffected by changes in $minSup$ values but exhibits non-uniform variation in alignment with the size of DT applications ($Size(A)$) due to dynamic allocation and deallocation of intermediate data structures, multi-stage processing demands, and periodic memory fragmentation and garbage collection, all of which cause temporary rises and falls in usage.

\subsection{Comparison}
\begin{figure*}[!htbp]
\centering	
\subfigure[Fault prediction accuracy]{
\includegraphics[width=0.28\linewidth, scale=8]{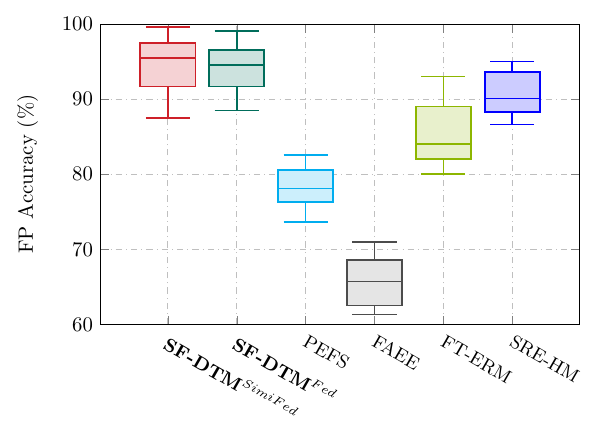}  
}
\subfigure[Resource contention]{\includegraphics[width=0.28\linewidth, scale=8]{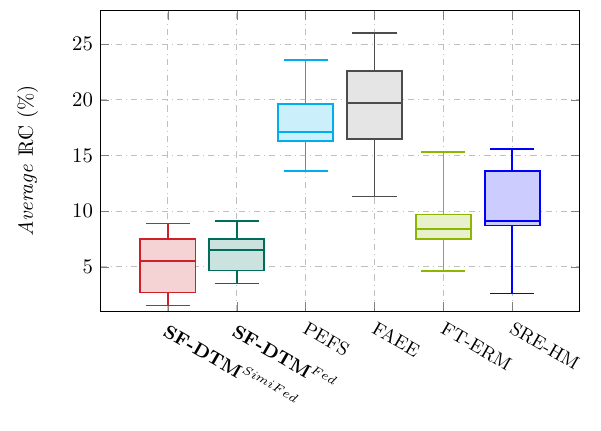}
}
\subfigure[Mean squared error ]{
  \includegraphics[width=0.28\linewidth, scale=8]{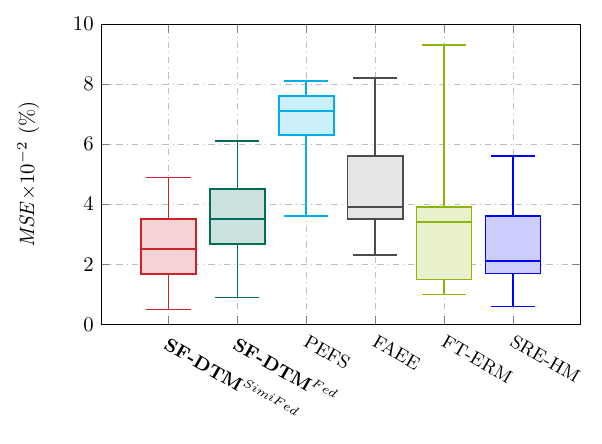}
}
\caption{Fault estimation comparison}
%\caption{Resource usage estimation errors}
\label{fig:comparison_fault_prediction}
\end{figure*}

The proposed SF-DTM model (referred to as SF-DTM$^{SimiFed}$) is compared against the baseline work SF-DTM$^{Fed}$, which represents one variant of our proposed model utilizing Federated learning. Additionally, we compare it with PEFS \cite{marahatta2020pefs}, FAEE \cite{sharma2019failure}, FT-ERM \cite{saxena2022fault}, and SRE-HM \cite{saxena2022high}. Figs. \ref{fig:comparison_fault_prediction}(a), \ref{fig:comparison_fault_prediction}(b), and \ref{fig:comparison_fault_prediction}(c) present a comparative analysis of fault prediction accuracy (\%), average resource contention (\%), and mean squared error, respectively, across all the aforementioned models for DT application size of 10 over an execution period of 400 minutes. It is observed that SF-DTM$^{SimiFed}$ consistently outperforms other models. This superior performance is due to its unique approach of integrating collaborative learning with the most similar group of local models for constructing a fault forecasting unit. This enhancement not only boosts performance but also enables more effective collaborative training of cloud-based DT applications, which is a capability previously unsupported by existing methodologies.

\begin{figure*}[!htbp]
\centering	
\subfigure[MTBF]{
\includegraphics[width=0.28\linewidth, scale=8]{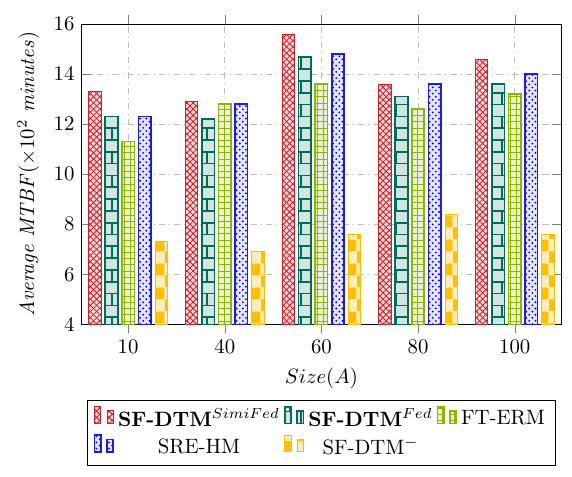}  
}
\subfigure[MTTR]{\includegraphics[width=0.28\linewidth, scale=8]{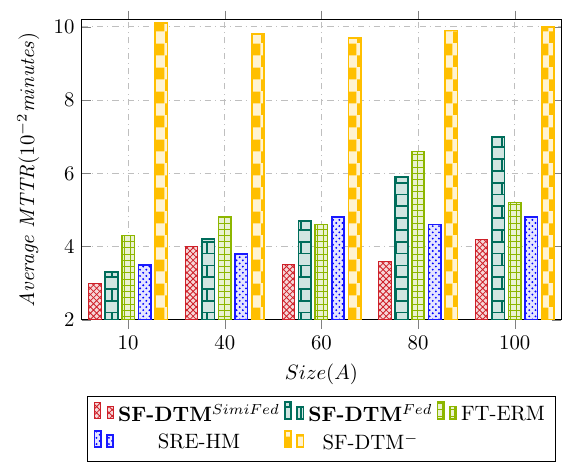}
}
\subfigure[Availability ]{
  \includegraphics[width=0.28\linewidth, scale=8]{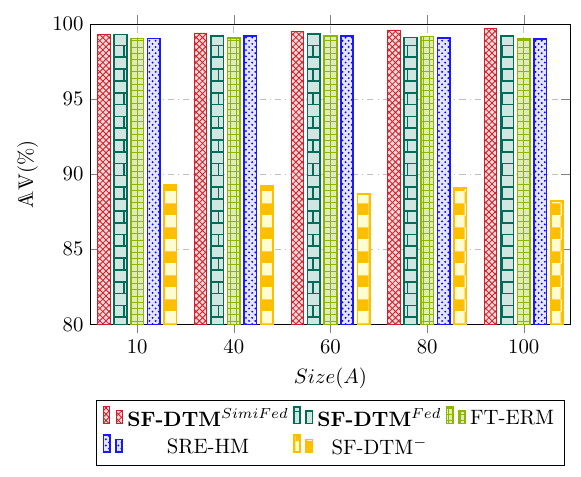}
}
\caption{Efficiency comparison for Self-healing and Fault-tolerance}
%\caption{Resource usage estimation errors}
\label{fig:mtbf_mttr}
\end{figure*}

\begin{figure*} [!htbp]
    \centering
    \includegraphics[width=0.7\linewidth]{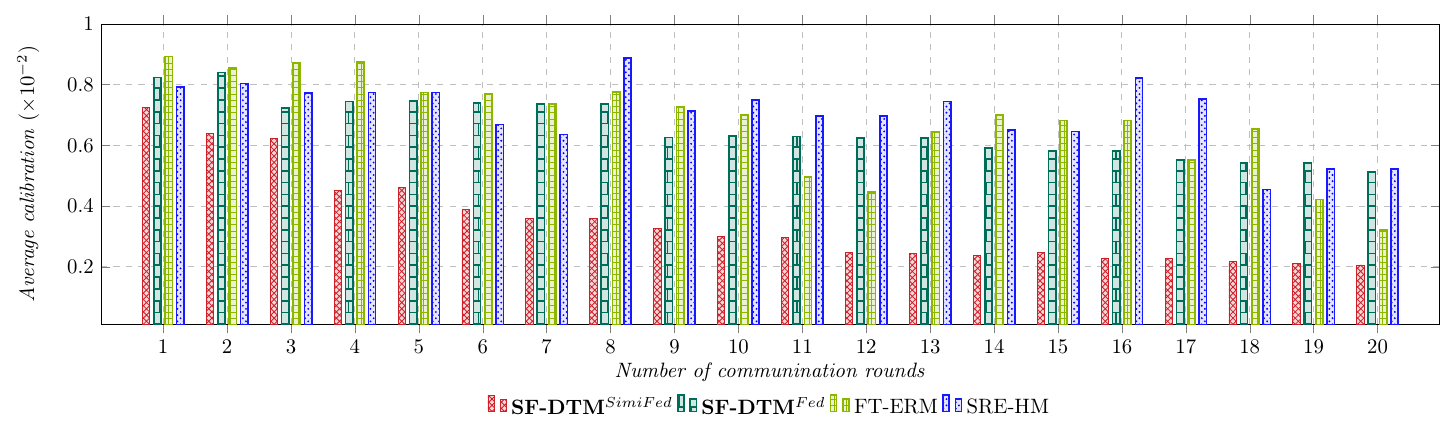}
    \caption{Comparative average calibration observed over consecutive number of retraining periods}
    \label{fig:calibration_comparison}
\end{figure*}

In Fig. \ref{fig:mtbf_mttr}, we compare the self-healing and fault-tolerance efficiency of the SF-DTM$^{SimiFed}$ model with SF-DTM$^{Fed}$, FT-ERM \cite{saxena2022fault}, and SRE-HM \cite{saxena2022high} across varying sizes of DT applications. Fig. \ref{fig:mtbf_mttr}(a) and Fig. \ref{fig:mtbf_mttr}(b) depict the increase in MTBF and decrease in MTTR in the following order: $SF-DTM^{SimiFed} > SF-DTM^{Fed} \ge SRE-HM > FT-ERM$. Furthermore, Fig. \ref{fig:mtbf_mttr}(c) reports the corresponding availability comparison, revealing that the availability is highest for $SF-DTM^{SimiFed}$. It outperforms SF-DTM$^{Fed}$, SRE-HM, FT-ERM and without SF-DTM ($SF-DTM^{-}$) by 0.0033\%, 0.0052\%, 0.0068\%, and 
 13.2\% respectively. The observed improvement is due to the precise fault estimation enabled by the collaborative SimiFed forecasting approach. Additionally, fault pattern learning and analysis strategies contribute to this enhancement, with potential for further scaling efficiency in DT applications. Fig. \ref{fig:calibration_comparison} reports a comparison of the average calibration observed during the twenty consecutive communication rounds  of the SF-DTM$^{SimiFed}$, SF-DTM$^{Fed}$, SRE-HM, and FT-ERM models for real-time processing in a DT application of size 10 over consecutive 20 retraining periods. It is evident that the SF-DTM$^{SimiFed}$ and SF-DTM$^{Fed}$ models demonstrate continuous improvement throughout the communication rounds and retraining intervals. Their average calibration errors decrease adaptively without accumulating. In contrast, the SRE-HM and FT-ERM models exhibit mild fluctuations and occasional accumulation of errors, likely due to the limitations  in their adaptive optimization and training methods when dealing with real-time, changing data.

\section{Conclusion} \label{conclusion}
This paper proposed the SF-DTM model, a pioneering solution for managing cloud-based DT applications. By combining collaborative resource estimation, Federated Learning with cosine Similarity integration (SimiFed), and self-healing fault-tolerance mechanisms, SF-DTM addresses critical challenges faced by cloud platforms. Through novel fault pattern learning and analysis, SF-DTM significantly improves availability, MTBF, and MTTR compared to existing approaches, as demonstrated through rigorous implementation and evaluation with real traces. The future work will focus on enhancing  fault-tolerance mechanisms and scalability of SF-DTM to accommodate complex and dynamic DT environments, as well as exploring applications beyond manufacturing and processing sectors to fully leverage its potential across diverse domains.

\bibliographystyle{IEEEtran}
	\bibliography{bibfile}
\begin{IEEEbiography}[{\includegraphics[width=1in,height=1.25in,clip,keepaspectratio]{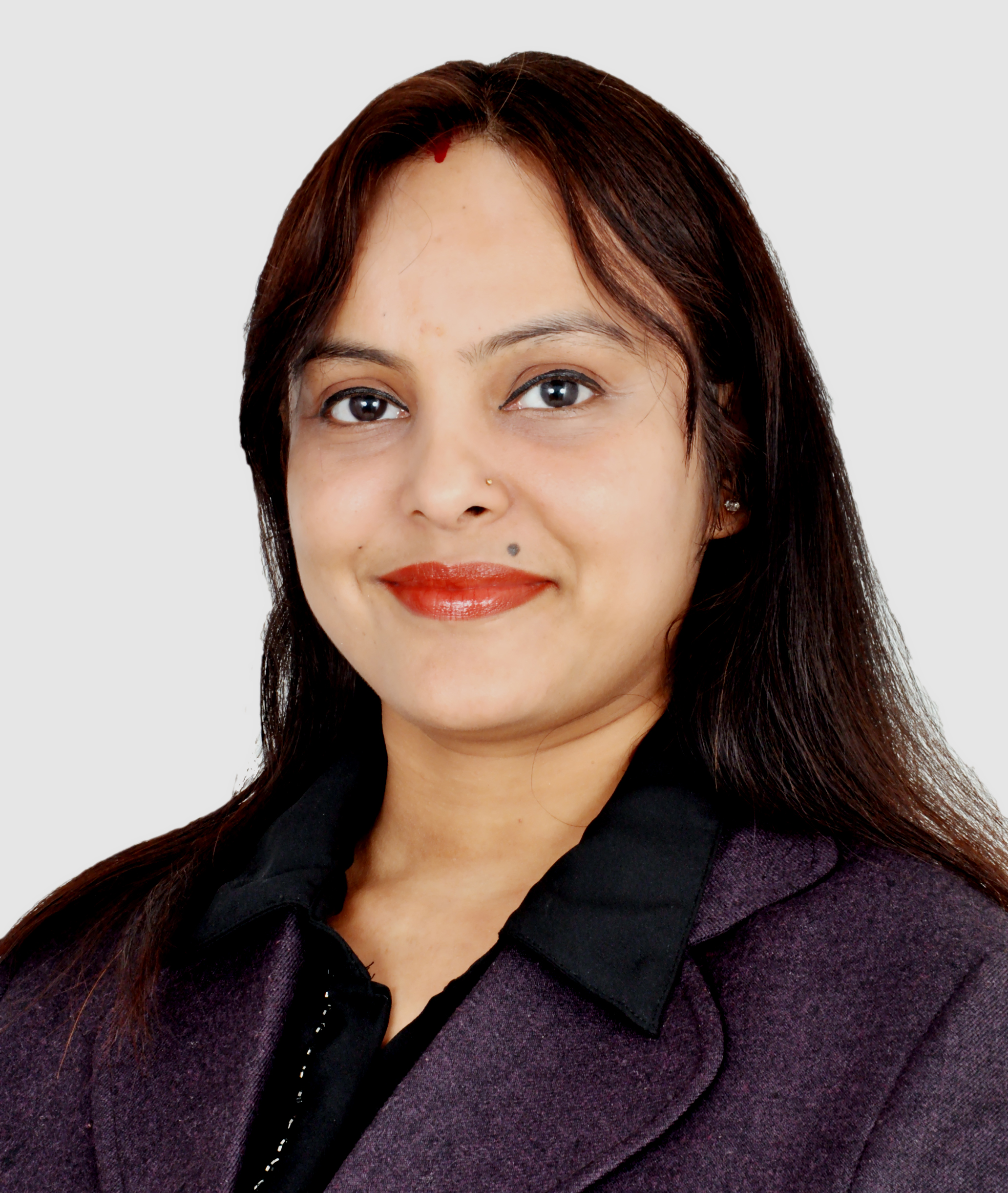}}]{Deepika Saxena} (Member, IEEE) is an Associate Professor in the Division of Information Systems at the University of Aizu, Japan. She received her Ph.D. from NIT Kurukshetra, India, and completed her postdoctoral research at Goethe University, Germany. She has received several prestigious awards, including the IEEE TCSC Early Career Researcher Award 2024, IEEE TCSC Outstanding Ph.D. Dissertation Award 2023, EUROSIM Best Ph.D. Thesis Award 2023, and the IEEE Computer Society Best Paper Award 2022. She is also a recipient of the JSPS KAKENHI Early Career Young Scientist Research Grant FY2024. Her research interests span neural networks, evolutionary algorithms, cloud resource management and security, traffic management, quantum machine learning, data lakes, and dynamic caching.
\end{IEEEbiography} 

\begin{IEEEbiography}[{\includegraphics[width=1in,height=1.25in,clip,keepaspectratio]{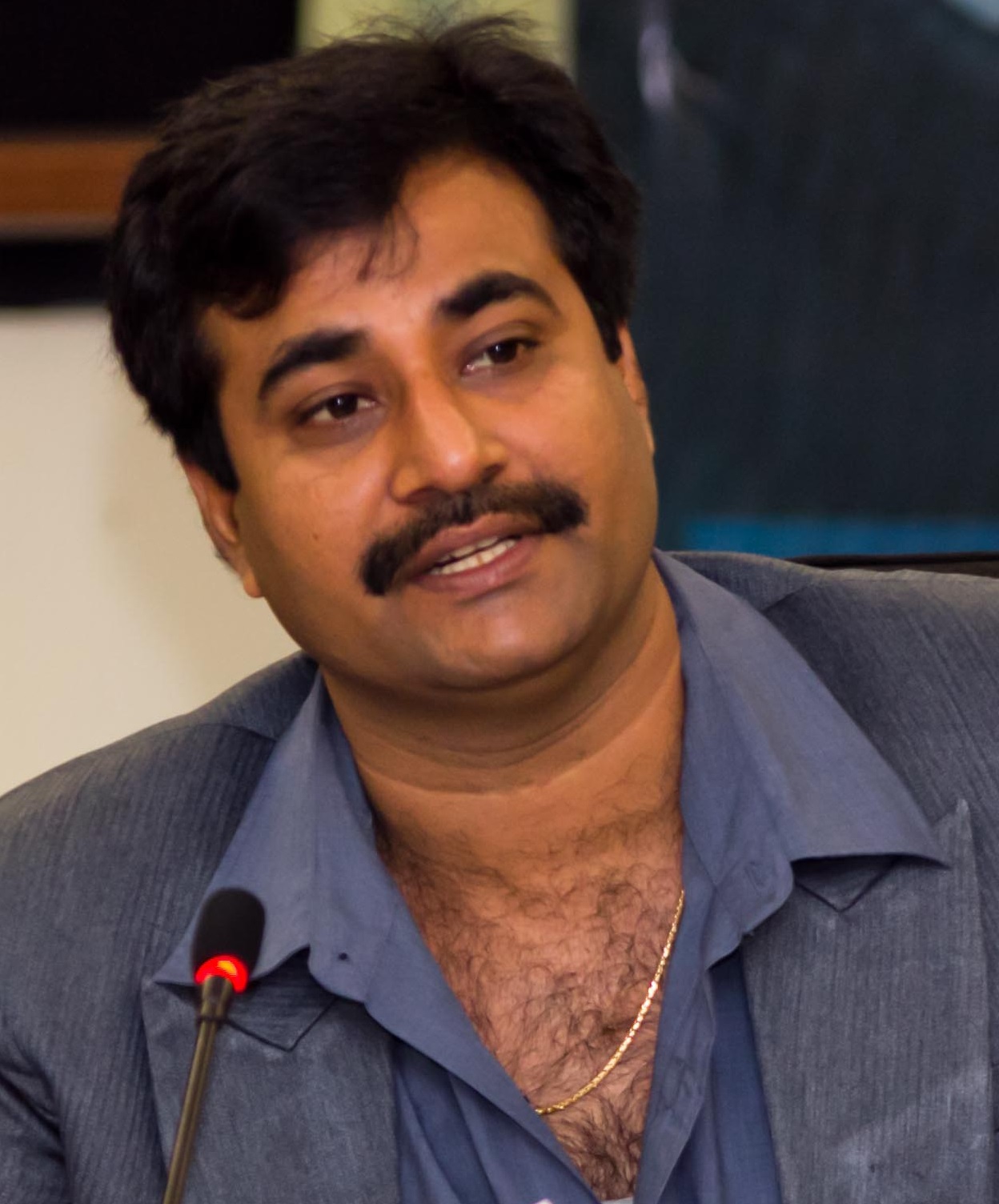}}]{Ashutosh Kumar Singh} (Senior Member, IEEE)  is a Professor and Director at IIIT Bhopal, India, and an Adjunct Professor at the University of Economics and Human Sciences, Warsaw, Poland. He earned his Ph.D. from IIT BHU, India, and completed his postdoctoral research at the University of Bristol, UK. With extensive research and teaching experience across India, the UK, and Malaysia, his expertise spans digital circuit design and testing, data science, cloud computing, machine learning, optimization algorithms, and security.  He has published over 400 high-impact papers in top journals, including IEEE TPAMI, TSC, TC, TSMC, TPDS, TII, TCC, FGCS, and Neurocomputing, etc. His IEEE Transactions on Cloud Computing paper received the IEEE Computer Society Best Paper Award 2022.

\end{IEEEbiography}

\end{document}